# Quantum multiple scattering

Robin Kaiser*

*Institut Non Linéaire de Nice, CNRS, Université de Nice Sophia-Antipolis, 06560 Valbonne, France*

The quest for Anderson localization of light is at the center of many experimental and theoretical activities. Atomic vapors play a particular role in this research field, as they show a number of specific properties that make them quite different from other materials used to look for Anderson localization. The very narrow resonance of the atomic line, the mechanical effects of the light on the atoms and the potential for quantum features of these scatterers call for more detailed analysis of the behavior of light in large and dense samples of cold atoms.

## 1. Introduction

The search for Anderson localization of non-interacting waves in three dimensions has been a dynamic field of research. Indeed, even more than 50 years after the seminal work by P. Anderson [1], very few experimental observations approach the ideal situation and there is a significant increase in the variety of set-ups used to approach this phase transition. In classical optics, experiments with semi-conductor powders were performed in 1997 [2], but effects related to spurious absorption have been discussed [3,4]. More recently, time-resolved experiments using 'white paint' (TiO2) have been performed [5] resolving to some extent the problem of absorption. Other waves have been used as well and the most complete set of experiments to date have probably been performed in acoustics [6]. A different approach has been used with matter waves in three dimensions, mapping the three-dimensional Anderson localization on a multifrequency kicked rotator [7]. As all these experiments are very delicate, one needs to be careful and only subsequent systematic work on all those systems might reveal which one is the closest to the ideal disorder-induced phase transition. It is maybe a good idea to recall that claims to see important but expected results should be subject to scrutiny by independent groups (remember the issue of cold fusion). In addition, a device that can observe the disorder-induced phase transition should be used to perform further experiments, only possible in this new phase. An example of such an approach is given by the realization of Bose–Einstein condensation (BEC) with dilute atomic gases. The first condensation signal in these systems [8] has been very convincing, but one could argue that the experimental 'smoking gun' has merely been consistent with the BEC transition. However, there has been such an overwhelming amount of experiments performed with BECs, that there is no doubt that the observation in [8] has indeed been the observation of the BEC transition. The Nobel prize in physics to pioneers in that field has been awarded after those confirmations. In the context of Anderson localization, one thus would like to have the same systematic approach of a double check by independent groups and new experiments in the localization regime.

## 2. How to trap a photon

Cold atoms have been used in the context of Anderson localization to approach the localization transition [9,10]. As such a system has very specific properties, a lot of 'homework' needed to be done in order to understand all the features of this unique material. In the past 10 years, mainly weak localization (more precisely coherent backscattering) has been studied with laser cooled samples and a number of review papers describe the work performed in this context [11,12,13,14]. In this article, we thus do not want to describe coherent backscattering, but rather focus on open questions that need to be addressed in the future. The general question to be addressed is how to trap a photon with many atoms. The underlying mechanisms to do so can be very different, but experimental signatures might be similar. For instance, in time-dependent experiments one might observe a long time decay and one wants to relate this decay to physical process involved in the atom-light interaction. In Figure 1 we show a possible map of different regimes to increase the lifetime of a photon in an atomic sample.

*Email: robin.kaiser@inln.cnrs.fr

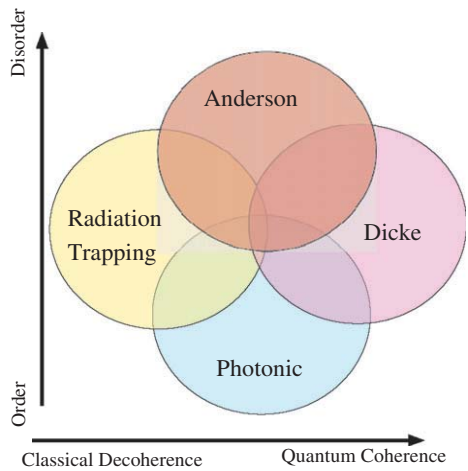

Figure 1. Various trapping mechanisms for a photon in a cloud of atoms. (The color version of this figure is included in the online version of the journal.)

One possibility to trap a photon is to use a well-ordered system where it would be possible to reduce the density of states as in photonic crystals. Such a regime has not yet been observed in experiments with atomic samples [15,16] and even in one dimension no firm experimental evidence has been obtained [17]. However, even though a complete bandgap seems difficult to reach with atomic samples, short or long range order might help other trapping mechanisms [18]. It seems therefore, a promising route to induce as much order as possible (in 1, 2 or 3 dimensions) when looking for how to trap a photon in atomic vapors.

If one only considers disordered media, one 'trivial' way to increase the time a photon spends in the system is to increase the size of the system. This regime, which we call the radiation trapping regime, has been well studied in the past, with hot atomic vapors [19,20] as well as with cold atoms [21,22,23]. In general, one might consider that radiation trapping in hot atomic vapors is based on an incoherent process, as Doppler (or collisional) broadening is dominant in such systems [19,20]. However, experiments in hot mercury vapors have identified evidence of coherent multiple scattering in the radiation trapping regime [24]. Unfortunately, these experiments are to some extent forgotten and no present research activity is being pursued with such systems (probably due to the wavelength in the UV needed for these experiments).

Another mechanism to obtain long lived states is to prepare subradiant Dicke states [25]. Despite some initial confusion, it has been quickly recognized that Dicke subradiance is based on a different physical mechanism than radiation trapping. Dicke superradiance has been observed in many different systems since the first observation [26], but only one experiment has reported a subradiant signature [38]. Indeed, it seems more difficult to isolate the long lived subradiant state from coupling to the short lived superradiant state. Dicke superradiance is easily observed in samples with either a small extent compared with the wavelength or in pencil shape configurations. A vast amount has been written on superradiance (see e.g. [28] for a recent report with updated references) and the role of quantum fluctuations in super- and sub-radiance might make the study of localization of photons in atomic vapors a very special way of trapping a photon.

Anderson localization of photons is yet another mechanism, based on disorder, to obtain localized, long-lived photonic states in an atomic sample. As atoms can to some extent be seen as highly efficient scatterers, a dense disordered sample of atoms has appealing properties to look for Anderson localization of light. Several experiments using 'classical' scatterers have reported signatures of Anderson localization of light in a three-dimensional disordered system [2,3,4,5] and, as mentioned in the introduction, further follow-up experiments will teach us more about the fascinating features of these systems. Experiments with atomic vapors have not yet shown evidence of Anderson localization of light. Even if no proof exists, it seems that hot atomic vapors will suffer from either collisional or Doppler broadening, which affect the coherence of the scattered light. On the other hand, if one considers Thomson scattering on charged particles, some dense astrophysical objects could become interesting systems to be studied. For cold atomic samples, where it seems more reasonable to neglect Doppler broadening and collisions, the main limitation comes from the densities one is able to obtain in a magneto-optical trap (MOT). A rough guide towards localization can be obtained with the Ioffe–Regel criterium [32], which reads: $kl \sim 1$, where $k$ is the wavevector of the light and $l$ the mean free path of the photon. The mean free path in the dilute regime scales as $1/\rho$, where $\rho$ is the atomic density. As in a standard MOT, one has $kl \sim 1000$, and one needs to increase the density by three orders of magnitude to reach the expected localization threshold. Current efforts in Nice are in progress and the use of quasi-resonant compressible dipole traps will be tried to reach the high density regime. Another route has been successfully followed in the group of M. Havey, using a far detuned, so-called QUEST trap to reach the high densities required for Anderson localization, and experiments to observe Anderson localization of light are in progress [29]. The required densities for rubidium atoms, for example, are of the order of $10^{13} - 10^{14} at/cc$. Such densities are obtained in a number of set-ups where BEC has been reached. However, the relatively low number of atoms and the bad duty cycle for such systems make it extremely challenging to look for photon localization in BECs.

As discussed above, there are a number of mechanisms that might trap a photon in a cloud of atoms and it will be interesting to see which mechanism will be the first one to efficiently trap a photon and, moreover, to understand how the mechanisms might influence or interact with each other. For instance, one might consider the Dicke subradiance trapping mechanism as a cooperative mechanism, which requires all atoms to be synchronized. Anderson localization on the other hand will prevent the spreading of the photon to the whole system and might thus prevent synchronization. On the contrary, one might consider that Dicke subradiance leads to long distance entanglement between atoms and the scattering of a photon can no longer be considered to be a result of a local dielectric constant. This situation is rather similar to the scattering on a non-local potential and most of the theoretical treatment of Anderson localization needs to be reconsidered in that case. In this respect, one might also reconsider the situation of Anderson localization of electrons at low temperature, where the disordered scattering matrix could be in a quantum fluid regime, with long range coherence. As Dicke super- and sub-radiance was first studied in small samples, whereas Anderson localization can only be a good description in large systems, we propose in Figure 2 a different map, which might divide the various regimes of how to trap a photon. This map is only a guide to stimulate discussions and its separation into different regimes of interest and should not be considered as a phase diagram. It only indicates that disorder, system size and number of atoms are related and cannot be changed independently. The diagonal lines in Figure 2 correspond to lines of constant atom number. Increasing the size of the sample at constant atom number thus allows us to cross from the Dicke regime to the radiation trapping regime, passing through the Anderson regime for large enough atom number.

A first approach to treat the localization of photons in a disordered medium while taking into account the quantum aspect of the scatterers has been described in [30]. The results of this first work points towards the dominant role of the cooperativity (Dicke states) over the disorder (Anderson). Further work along these quantum aspects of localization will be needed in order

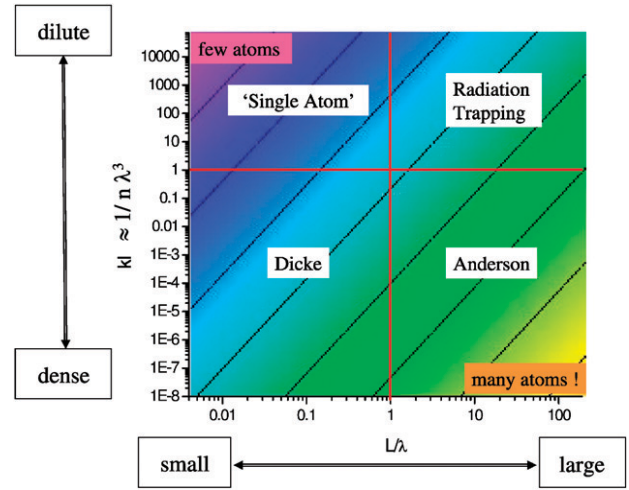

Figure 2. Classical and quantum trapping of photons. (The color version of this figure is included in the online version of the journal.)

to understand the possibilities of how to trap a photon in a cloud of atoms.

### 3. Optical response of dense atomic vapors

Let us now turn to a slightly more quantitative description of the optical response of a dense cloud of cold atoms. A simple approach using a semiclassical description of the atoms can be used to derive many qualitative features of the optical response in dense atomic vapors. Following previous work [31], we will show that the resonant mean free path will be affected when approaching the high density limit and that a collective red shift breaks the red/blue symmetry existing in dilute samples.

As the 'Holy Grail' is the observation of strong localisation of light, we will start by a simple Ioffe–Regel criterion for strong localization:

$$kl \sim 1. \qquad (1)$$

We would like to stress that this criterion is mainly a rough guide of when to look for strong localization type of effects and cannot be considered as a prediction of the threshold. Important prefactors could appear and make the threshold change by a relevant amount. However, our purpose here is to discuss possible qualitative effects when studying the optical response of dense atomic samples.

When using resonant scatterers such as two-level atoms, the cross-section can be of the order of $\sigma_{res} \sim 3\lambda^2/(2\pi)$ and strong localization might be expected for $\rho\lambda^3 \sim 1$. Note that in this result, only the optical wavelength is relevant, in contrast to the threshold for Bose–Einstein condensation where one requires $\rho \lambda_{dB}^3 > \sim 2.613$ and where the De Broglie wavelength is the important parameter. The straightforward route towards Anderson localization thus seems a brute force increase of spatial density. In this case, however, collective radiation effects might become too important to be neglected and could yield a larger mean free path than for independent scatterers. Indeed, consider for example powder of amorphous $SiO_2$. For large air spacings between the different grains of $SiO_2$, the sample will be diffusive and a photon mean free path scales as $1/\rho_{SiO_2}$ where $\rho_{SiO_2}$ is the density of grains. However, in the extreme limit of compact $SiO_2$, we recover transparent glass. It is thus clear that when increasing the spatial density of scatterers, the photon mean free path will pass through a minimum.

Let us estimate this correction of the photon mean free path due to the correlation arising from recurrent scattering between atoms. This can also be interpreted as dipole–dipole Van der Waals effects. Indeed, in dense media, the polarizability $\chi$ is modified due to the local field effect (Lorentz–Lorenz formula). Neglecting the long range correction, also known as Purcell effect [35] we have:

$$\chi = \frac{\rho \alpha}{1 - \frac{1}{3}\rho \alpha} \quad (2)$$

and for corrected dielectric constant $\varepsilon = 1 + 4\pi \chi$ we have: $\frac{\varepsilon - 1}{\varepsilon + 2} = \frac{1}{3}\rho\alpha$. We use $\alpha = \frac{1}{\delta - i\frac{\Gamma}{2}}\frac{\alpha_0}{2}\omega_L$ and $\frac{3\lambda^2}{2\pi} = \alpha_0 k \frac{\omega_{at}}{\Gamma}$, where the atomic parameters are the wavelength $\lambda = 2\pi/k = 2\pi c/\omega_{at}$, the excited state width $\Gamma$, the static polarizability $\alpha_0$ and the laser-atom detuning $\delta = \omega_L - \omega_{at}$. One finally obtains for the corrected dielectric constant $\varepsilon$:

$$\varepsilon = 1 + 4\pi\rho \frac{\delta - \Delta\omega_{LL} + i\frac{\Gamma}{2}}{(\delta - \Delta\omega_{LL})^2 + \frac{\Gamma^2}{4}} \frac{\alpha_0}{2}\omega_L \quad (3)$$

where the collective shift, which can be seen as the Lorentz–Lorenz shift $\Delta\omega_{LL}$, is given by:

$$\Delta\omega_{LL} = -\frac{\rho\lambda^3}{8\pi^2}\Gamma. \quad (4)$$

This red-shift of the resonance is thus expected to be small for dilute samples, whereas for samples close to the Ioffe–Regel criterion this shift can become substantial. We thus have a density-dependent frequency shift, closely resembling the expression of the collective Lamb shift [36]. The above derivation on the shift of the center of line is of course very simplified and a more rigorous approach can, for example, be found in [33,34]. However, this simplified approach for the line shift, allows us to easily make some qualitative predictions on the consequence of the Lorentz–Lorenz shift on the threshold of Anderson localization. Consider for instance the resonant condition ($\delta = 0$).

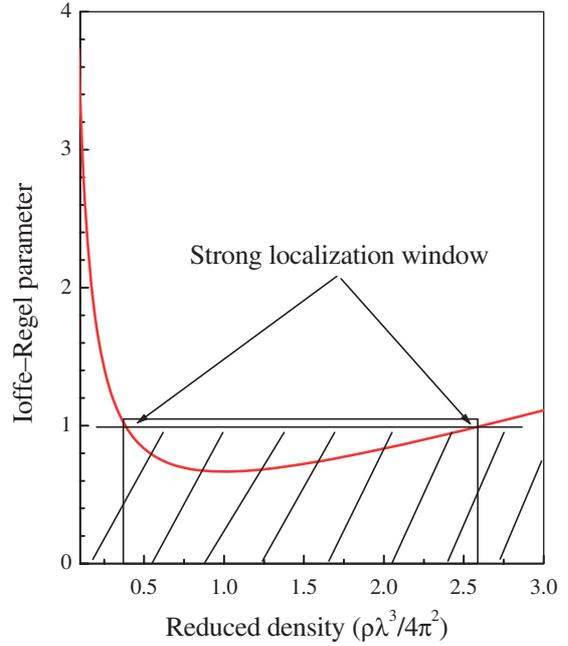

Figure 3. Modified Ioffe–Regel criterium for resonant excitation. (The color version of this figure is included in the online version of the journal.)

In this case the imaginary part of the atomic polarizability is reduced by a factor of $\frac{\frac{\Gamma^2}{4}}{(\Delta\omega_{LL})^2 + \frac{\Gamma^2}{4}} = \frac{1}{\left(\frac{\rho\lambda^3}{4\pi^2}\right)^2 + 1}$. This is equivalent to a reduced resonant cross-section: $\widetilde{\sigma}_{res} = \sigma_{res}\frac{1}{\left(\frac{\rho\lambda^3}{4\pi^2}\right)^2 + 1}$ corresponding to an increased 'on-resonant' mean free path: $\widetilde{l}_{res} = \frac{1}{\rho\sigma_{res}} = \frac{\left(\frac{\rho\lambda^3}{4\pi^2}\right)^2 + 1}{\rho\frac{3\lambda^2}{2\pi}}$ and one obtains the Ioffe–Regel criterium (Figure 3)

$$k\widetilde{l}_{res} = \frac{1}{3}\frac{\left(\frac{\rho\lambda^3}{4\pi^2}\right)^2 + 1}{\frac{\rho\lambda^3}{4\pi^2}} \quad (5)$$

which is minimal for $\frac{\rho\lambda^3}{4\pi^2} = 1$ and then takes the minimum value:

$$k\widetilde{l}_{min} \simeq \frac{2}{3} < 1. \quad (6)$$

This model seems to predict that strong localisation of resonant light in dense cold atomic vapors can only be obtained in a narrow window of densities (for $\frac{3-\sqrt{5}}{2} < \frac{\rho\lambda^3}{4\pi^2} < \frac{3+\sqrt{5}}{2}$). It is, however, possible to compensate the shift of the resonance by taking:

$$\delta + \Delta\omega_{LL} = 0. \quad (7)$$

Indeed, the detuning dependent Ioffe–Regel parameter is given by

$$k\widetilde{l}_{res} = \frac{1}{3}\frac{\left(\frac{2\delta}{\Gamma} + \frac{\rho\lambda^3}{4\pi^2}\right)^2 + 1}{\frac{\rho\lambda^3}{4\pi^2}}. \quad (8)$$

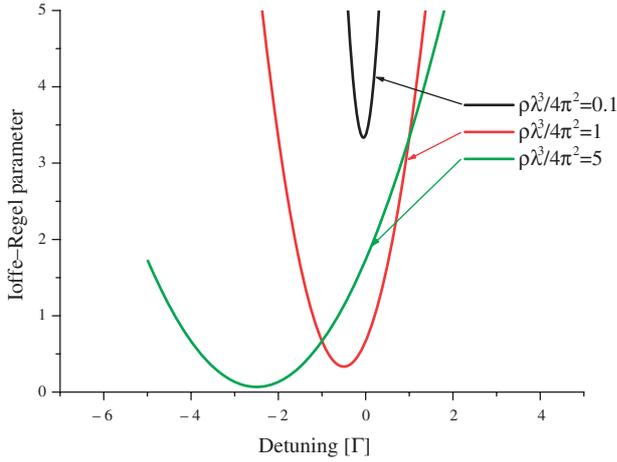

Figure 4. Ioffe–Regel criterium for detuned excitation. (The color version of this figure is included in the online version of the journal.)

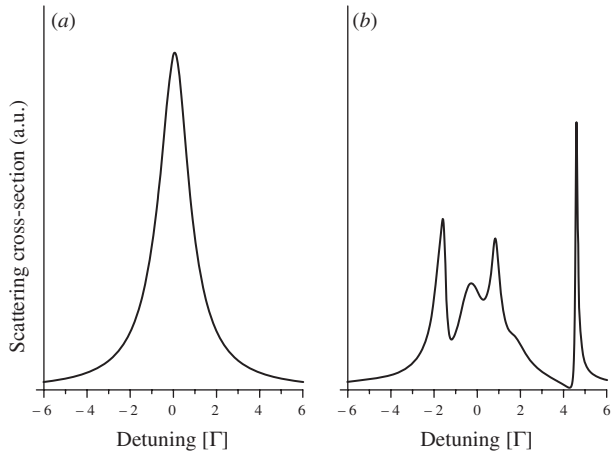

Figure 5. Spectra for a dilute cloud corresponding to $kl = 10$ (a) and a dense cloud corresponding to $kl = 0.1$ (b) of $N_{at} = 10$ cold atoms.

This should compensate for this Lorentz–Lorenz shift and it should then be possible to keep the same cross-section as without local field effects. Figure 4 shows the shift of the optimum detuning for increased densities.

At the optimum detuning, the Ioffe–Regel parameter is given by

$$k\widetilde{l_{opt}} = \frac{4\pi^2}{3\rho\lambda^3}. \quad (9)$$

The constraint on the atomic density thus seems to be the same as if one had neglected the Lorentz–Lorenz shift. One should, however, be careful of taking this rescaled threshold condition in a too quantitative way. Indeed, the Lorentz–Lorenz shift is an average effect.

The real optical response of a large cloud of cold atoms does contain a large number of narrow and broad resonances. The average shift thus corresponds to an inhomogeneous line broadening and shift. In Figure 5 we show two spectra (obtained from an effective Hamiltonian [30]) for a dilute cloud (a) and a dense cloud (b) of cold atoms, illustrating that the collective response of the atoms is not completely described by a shifted Lorentzian.

## 4. Nonlinear optics and Anderson localization

It is important to mention at this stage that a huge amount of literature exists on nonlinear optics in dense atomic vapors. In particular, many pump/probe schemes, involving more than a single low intensity laser beam allow for a large variety of effects to appear. Spectacular results have thus been obtained in so-called slow light experiments, often based on an electromagnetically induced transparency. In addition, nonlinear optics at extremely low levels of the probe intensity have been achieved [37].

However, in the context of Anderson localization, it is important to notice that it is crucial to distinguish effects due to large spatial densities, which can be described with a continuous density profile from effects arising from fluctuations around such a large smooth density distribution. Nearest neighbor contributions can in such limits not be neglected as already noted in the context of cold Rydberg atoms [38]. Even though experimental efforts are still some way from a clear identification of Anderson localization with cold atoms, one can speculate what will happen once the threshold is crossed.

One important consequence of strongly localized modes of photons in a cloud of cold atoms is that the local field will be immensely enhanced. One clearly expects important enhancement for localized modes, similar to what happens in a high quality Fabry–Perot-like cavity. However, in the case of strong localization with cold atoms, this enhancement could reach astonishing values. A precise evaluation of the enhancement factor depends on the localization length $\xi$, which gives the length scale over which the electromagnetic wave decreases exponentially. Let us assume that the localization length is of the order of the optical wavelength $\lambda$. Close to the threshold one expects to have one atom in a volume of the order of $\lambda^3$. As each atom cannot absorb more than one photon, one will be in the very interesting regime of an optical blockade of photons. If there are $N$ atoms in the localization volume, then this blockade will appear at a photon number of $N$. This is clearly an interesting regime to explore, as one enters the cavity QED regime in the absence of any external cavity.

But even when one does not enter the blockade regime, strong saturation effects will appear. Indeed, for a localization $\xi$, the energy density $W_{1h\nu} = \frac{1}{2}\varepsilon_0 E^2$ of one single localized photon will be:

$$W_{1h\nu} = \frac{h\nu}{\xi^3}. \tag{10}$$

On the other hand, we know that the atomic excitation saturates for values of $E$ corresponding to: $\frac{1}{2}\varepsilon_0 c E^2 = I_{sat} = \frac{\pi}{3}\frac{hc\Gamma}{\lambda^3}$. This consideration leads to strong saturation and nonlinear response of the cloud for a localization length of

$$\xi \sim \lambda \left(\frac{\omega}{\Gamma}\right)^{1/3} \tag{11}$$

which for rubidium atoms, for example, is of the order of $\xi \sim 400\lambda$. So even for such large localization lengths the atomic response will become nonlinear for a single localized photon! Such high nonlinearities will be good candidates for optical bistability and the connection of Anderson localization and the so-called instrinsic optical bistability [39,40] merits further studies.

## 5. Conclusion

We have discussed in this article how the study of Anderson localization with cold atoms is related to a number of other interesting fields of research: radiation trapping, Dicke super- and sub-radiance and photonic crystals. The interplay between these mechanisms needs further study but it is important to start by recognizing the potential impact of competing effects when studying the localization of photons. We have also given some estimation of collective shifts and how they can influence the threshold of Anderson localization. Nonlinear effects are clearly expected close to and in the localization regime and one would need to include even further complexity in this context, such as the mechanical effects of light on the atoms. When red detuned light is used (taking into account the Lorentz–Lorenz shift) one expects atoms to be attracted towards the localized modes. This can change the spatial density and thus shift the resonance, which is yet another mechanism to induce bistability in such a fascinating system.

## Acknowledgements

It is a pleasure to thank my colleagues in Nice as well as E. Akkermans, Ph. Courteille, M. Havey, T. Gallagher, D. Gauthier and P. Pillet for many fruitful discussions on topics presented in this article. Financial support from is ANR project ANR-06-BLAN-0096 is also acknowledged.